# A Constructive Theory of Closed-Form Rate–Distortion Functions

Dazhuan Xu[1,2], Yongming Huang[1,3], and Xiaohu You[1,3*]

[1] Purple Mountain Laboratories, Nanjing, China

[2] Nanjing University of Aeronautics and Astronautics, Nanjing, China

[3] Southeast University, Nanjing, China

## Abstract

Rate–distortion functions are defined by a variational problem, and analytic solutions depend on the relationship between the source distribution and the distortion measure. This paper develops a posterior-based construction framework with conditional entropy as a common distortion coordinate. Building on the rate–entropy formulation, we establish entropy-level attainability for discrete sources with finite entropy and absolutely continuous sources with integrable log densities. We also give a constructive rate–entropy coding proof for finite-alphabet sources and for such continuous sources, without a second-moment assumption. For physical distortion, marginal compatibility ensures that a candidate posterior and output distribution recover the prescribed source, while a Gibbs representation and a kernel mass bound certify attainment of maximum conditional entropy. The parameter structure of the negative log posterior identifies candidate fixed distortion functions. Excess rate equals a conditional entropy gap and decomposes into budget slack and posterior mismatch. Two construction paths, based on convolution decompositions and conjugate posteriors, recover several established analytic results, including the Gamma-source rate–distortion function under Itakura–Saito distortion. They also yield a closed-form parametric expression for zeta sources under divisibility-constrained logarithmic multiplicative distortion. For negative binomial sources under one-sided counting distortion and for Gamma sources under one-sided mean error, exact expressions are obtained on explicitly specified intervals. A residual-count reparametrization of the Poisson entropy budget, together with Gamma and Dirichlet counting channels, further connects model parameters to entropy budgets. The framework links distortion selection, compatible-channel construction, and optimality verification within a common analytic procedure.

## 1 Introduction

Rate–distortion theory provides the foundation for lossy source coding and data compression by characterizing the minimum information rate under a distortion constraint [1]–[3]. More than six decades after Shannon's formulation, closed-form solutions for specific sources still rely largely on particular combinations of source distributions and distortion measures. Even a continuous uniform source, whose density is elementary, presents a nontrivial rate–distortion problem under squared error. Its finite support does not make the optimization straightforward. The difficulty therefore involves the source distribution, the distortion function, and their structural relationship.

Information rate and distortion differ, first of all, in what they measure. Mutual information quantifies the source information retained by a representation, whereas mean squared error, symbol error rate, and other task losses have distinct units and mathematical forms. Moreover, the general definition of a distortion function permits many choices without specifying which choices support analytic solutions. Even after the distortion is fixed, finding the optimal test channel remains a variational problem; finite support, boundaries, and density discontinuities can complicate its solution. This motivates a joint study of how to identify suitable distortion measures and construct their optimal channels.

* This work was supported by the National Natural Science Foundation of China under Grant 62271254. (Corresponding author: Dazhuan Xu; e-mail: xudazhuan@nuaa.edu.cn.)

The authors' earlier work defined the rate–entropy function using conditional entropy as the distortion measure, providing a common information coordinate for relating task-dependent criteria to rate [4]. The present paper develops the connection between physical distortion and entropy distortion through posterior families compatible with a prescribed source. Terms coupling the source and representation variables in the negative log posterior identify candidate fixed distortions. Posterior entropy determines the rate, and conditions for attaining maximum conditional entropy certify optimality. These steps bring distortion selection and rate–distortion computation into a common analytic construction.

### 1.1 Related Work

Shannon established the rate–distortion function and its parametric formulation [1]. Berger gave a systematic treatment of the optimality conditions [2], and Csiszár studied extremum problems and test-function conditions on general alphabets [5]. Classical analytic results include Gaussian sources under squared error, Laplace sources under absolute error, exponential sources under one-sided error, and equiprobable discrete sources under Hamming distortion [2], [3], [6]. In these settings, maximum-entropy errors and independent backward channels yield equality in the corresponding Shannon lower bounds. Subsequent work has characterized how closely the rate–distortion function approaches the lower bound at low distortion and has extended the construction of such bounds [7], [8]. Blahut's rate–distortion iteration and Arimoto's related channel-capacity algorithm established classical numerical approaches [9], [10].

For vector Gaussian sources, the theory extends from total mean squared error to several covariance-based constraints. Under a trace constraint, reverse water-filling specifies the optimal error allocation among eigendirections. A matrix constraint directly bounds the error covariance and, under the relevant matrix-order conditions, yields a rate expressed through a ratio of covariance determinants [2], [3]. Marsiglietti and Kostina studied rate–distortion bounds under a determinant constraint; the relative-entropy gap bound in their Theorem 11, Eq. (61), vanishes for Gaussian sources and therefore determines the exact rate function [11]. Trace, matrix, and determinant budgets measure total error, directional error structure, and error volume, respectively. Their different feasible sets illustrate the dependence of analytic results on the distortion criterion even for the same source. Related work on remote vector Gaussian coding with decoder side information provides explicit expressions under covariance matrix constraints [12].

For positive sources, quotient-based distortions extend the analysis from differences to relative errors. Buzo et al. developed rate–distortion bounds for quotient-based costs, motivated by the gain term in linear predictive speech coding. Under Itakura–Saito (IS) distortion, they obtained the exact Gamma-source rate–distortion function and an optimal multiplicative backward channel, and studied a low-distortion approximation for optimal scalar quantization [13]. The Gamma–IS result is cited as a known solvable case in [14]. Watanabe and Ikeda studied Gamma-type sources under absolute-log distortion. A logarithmic transformation converts the problem to one with absolute-error distortion, leading to a parametric rate–distortion expression and a characterization of the continuous and discrete components of the optimal reconstruction distribution [15]. These results demonstrate the analytic value of relative errors and coordinate transformations. A further question is how the different solvable source–distortion pairs fit into a common construction method.

Logarithmic loss evaluates a probability distribution as the reconstruction and is suited to settings in which the decoder retains source uncertainty. For a given representation, the true posterior minimizes expected logarithmic loss, and the minimum equals conditional entropy. A distortion constraint thus becomes a constraint on remaining uncertainty. Courtade and Weissman used this relation to characterize the rate–distortion regions of the two-encoder source-coding problem and the CEO problem under logarithmic loss [16]. No and Weissman studied correspondences between lossy compression under logarithmic loss and under general distortion criteria [17]. Shkel and Verdú developed a single-shot analysis of lossy coding under logarithmic loss [18]. These studies establish probability-valued reconstruction as a useful way to relate retained information to compression rate.

The logarithmic-loss literature uses conditional entropy to characterize distortion constraints in specific coding models. The authors' earlier work took conditional entropy itself as the budget and defined the rate–entropy function [4]. When the target entropy level is attainable, the minimum information rate equals the source entropy minus that level. That work also related exponentiated entropy distortion to distributional scale parameters, obtaining closed-form rate–entropy expressions for uniform, Cauchy, chi-square, and vector Gaussian sources. Conditional entropy thereby provides both a measure of remaining uncertainty and a coordinate connecting physical scales to rate.

These developments supply complementary ingredients: classical optimality conditions certify channels, maximum-entropy distributions and independent decompositions construct exact solutions, and conditional entropy relates information rate to different error scales. This paper combines them into a reusable procedure: construct a compatible posterior family, identify a fixed distortion from its parameter structure, and verify optimality over an explicit parameter range. Recovering established results and deriving further analytic expressions provide concrete tests of the procedure.

### 1.2 Approach and Main Contributions

We study analytic rate–distortion construction through the posterior representation of a test channel. For a fixed source and distortion function, whenever the relevant entropies and optimal values are well defined, the mutual information identity converts rate minimization into conditional entropy maximization:

$$R_d(D) = H_\mu(X) - E_d(D).$$

Here $E_d(D)$ is the largest conditional entropy permitted by distortion budget $D$. On the attainable entropy interval, $R_e(E) = H_\mu(X) - E$, so the effect of the physical distortion is captured by $D \mapsto E_d(D)$. Starting from a compatible posterior family, we identify a structurally appropriate fixed distortion, verify attainment of maximum conditional entropy, and vary the parameter to obtain the rate–distortion curve and its valid range. When a task already specifies the distortion, the same conditions test the optimality of a candidate posterior.

The contributions are as follows.

1) Entropy attainability and coding. Building on the rate–entropy definition in [4], we allow distinct source and representation alphabets and relate physical distortion budgets, maximum conditional entropy, and minimum information rate. For discrete sources with finite entropy and absolutely continuous sources with integrable log densities, we establish attainability of general entropy levels and determine the linear region of the rate–entropy function. The continuous construction uses equal-probability partitions and time sharing without a second-moment assumption. We further give a constructive rate–entropy coding proof for finite-alphabet sources and the above continuous sources, specifying encoding regions, messages, and decoder posteriors that achieve the corresponding block entropy budgets.

2) Analytic construction through compatible posteriors. We represent a test channel by its induced output distribution and backward channel. Marginal compatibility ensures that their mixture realizes the prescribed source. In a posterior formulation of classical optimality conditions [2], [3], [5], a Gibbs form together with a kernel mass bound over the entire output alphabet certifies attainment of maximum conditional entropy. The parameter structure of the full negative log posterior identifies candidate fixed distortions, and parameter variation yields rate–distortion expressions over their admissible ranges. The excess rate of a general channel equals its conditional entropy gap and decomposes into budget slack and posterior mismatch. In residual coordinates, posterior mismatch further separates into residual–output dependence and residual distribution mismatch. This connects distortion selection, channel construction, and optimality verification.

3) Two construction paths and their analytic results. The convolution path uses independent residuals and transform product identities to ensure marginal compatibility. The conjugate-posterior path starts from a forward test channel and the source distribution, which determine a compatible joint distribution and posterior. The two paths encompass established Gaussian, Laplace, exponential, and Pareto examples and recover the exact Gamma–IS result [13]. For zeta sources, we derive a compatible output distribution and a complete parametric rate–distortion expression under divisibility-constrained logarithmic multiplicative distortion, using the Euler product to realize an integer product decomposition. For negative binomial sources under one-sided counting distortion, we obtain an exact expression in terms of the source entropy over a range determined by nonnegativity of the reconstruction probabilities; the geometric case covers the full distortion range. Further results include an exact interval for Gamma sources under one-sided mean error, an equivalent residual-count coordinate for Poisson entropy budgets, and rate–entropy parametrizations of Gamma and Dirichlet counting channels covering every finite attainable entropy budget. Beta sources are included as the two-dimensional Dirichlet case.

Conditional entropy connects physical distortion to information rate, while compatible posteriors connect probability structure to optimality. Results for fixed expected costs, physical-parameter coordinates of entropy budgets, and rate–entropy parametrizations of counting representations are stated with their respective constraints and ranges. Together, the recovered and derived results establish posterior structure as a common basis for selecting distortion measures and solving for optimal channels.

### 1.3 Organization of the Paper

Section 2 develops the rate–distortion and rate–entropy framework and proves entropy attainability. Section 3 gives sufficient conditions for attaining maximum conditional entropy, rate-gap identities, parametric representations, and conditions for identifying a fixed distortion from a posterior family. Section 4 develops convolution constructions, including negative binomial and zeta examples. Section 5 studies conjugate posteriors for relative-error and counting representations. Section 6 proves the rate–entropy coding theorem constructively. Section 7 presents conclusions and future directions. The appendix summarizes the principal distortion criteria, analytic results, and their sources.

## 2 A Unified Framework for Rate–Distortion and Rate–Entropy Functions

### 2.1 Distortion Between Source and Representation Alphabets

Let $X \in \mathcal{X}$ be a source random variable and let $Y \in \mathcal{Y}$ be the output of a test channel, referred to as a representation of the source. The source alphabet $\mathcal{X}$ and the representation alphabet $\mathcal{Y}$ are standard Borel spaces and may differ in dimension, support, and interpretation. A source distribution $P_X$ and a test channel $P_{Y|X}$ determine the joint distribution and the induced output distribution $P_Y$.

A distortion function is a measurable map $d: \mathcal{X} \times \mathcal{Y} \to [0, +\infty]$. Its value $d(x, y)$ specifies the cost of representing a source value $x$ by $y$, with average distortion constrained by $\mathbb{E}d(X, Y) \le D$. This cost may describe reconstruction error, classification error, or another task loss. The alphabets need not coincide, and the cost need not be symmetric. Logarithmic loss for continuous sources may take negative values; it is treated below under the stated integrability conditions.

To treat discrete and continuous sources in a common notation, fix a reference measure $\mu$ and write $p_X = dP_X/d\mu$. Define

$$H_\mu(X) = -\int p_X(x) \log p_X(x)\, \mu(dx), \qquad E = H_\mu(X|Y). \quad (2.1)$$

Counting measure yields Shannon entropy, whereas Lebesgue measure yields differential entropy. All logarithms are natural, and rates are measured in nats. Mutual information is defined through relative entropy. Entropy differences are used when the relevant log densities are integrable; additional finiteness conditions are stated where needed.

For a fixed source distribution, representation alphabet, and distortion function, the rate–distortion function is

$$R_d(D) = \inf_{P_{Y|X}:\, \mathbb{E}d(X,Y) \le D} I(X; Y). \quad (2.2)$$

The representation determines a posterior distribution, which may subsequently be used to produce a point reconstruction. These two roles are distinguished next.

## 2.2 Posterior Distributions and Point Reconstructions

Given $P_X$ and $P_{Y|X}$, the representation $Y$ determines the posterior distribution $P_{X|Y}$. As a distribution-valued statistic of $Y$, this posterior preserves all information in $Y$ about $X$ and describes the uncertainty remaining after encoding. It therefore provides the basis for subsequent inference and point reconstruction. The posterior description leaves the test channel and its distortion criterion unchanged: the average distortion remains $\mathbb{E}d(X,Y)$.

When a point reconstruction is required, the decoder produces $\hat{X}$ from the posterior. Deterministic choices include a maximum a posteriori (MAP) estimate, the posterior mean, and a posterior median, whenever these estimates are well defined. Posterior sampling instead draws $\hat{X}$ from $P_{X|Y}$ using decoder randomness independent of $(X,Y)$.

Each such reconstruction satisfies the Markov relation $X \to Y \to \hat{X}$, and hence $I(X;\hat{X}) \le I(X;Y)$. When the mutual information is finite, the information loss is

$$I(X;Y) - I(X;\hat{X}) = I(X;Y \mid \hat{X}) \quad (2.3)$$

The loss vanishes precisely when $\hat{X}$ retains all information in $Y$ about $X$, equivalently, when $X$ and $Y$ are conditionally independent given $\hat{X}$. The MAP estimate, the posterior mean, and posterior sampling do not in general guarantee this sufficiency property.

## 2.3 Entropy Distortion and the Rate–Entropy Function

The conditional entropy $E = H_\mu(X|Y)$ measures the uncertainty remaining after the representation is observed. It is also the minimum expected logarithmic loss for that representation, attained by the true posterior. Following the terminology of the authors' earlier work [4], define the rate–entropy function by

$$R_e(E) = \inf_{P_{Y|X}:\, H_\mu(X|Y) \le E} I(X;Y). \quad (2.4)$$

Here the representation alphabet and the test channel may be chosen separately for each entropy budget. Every feasible channel satisfies

$$I(X;Y) = H_\mu(X) - H_\mu(X|Y) \ge H_\mu(X) - E. \quad (2.5)$$

Together with the nonnegativity of mutual information, this gives $R_e(E) \ge [H_\mu(X) - E]^+$, where $[a]^+ = \max\{a, 0\}$. For a target level not exceeding the source entropy, a channel with conditional entropy exactly $E$ attains the linear lower bound. We next formulate this attainability requirement in terms of a backward channel.

## 2.4 Entropy Attainability and Time Sharing

**Definition 1 (Attainability of an entropy level).**

For a fixed source distribution $P_X$, an entropy level $E$ is attainable if there exist an output distribution $P_Y$ and a backward channel $K(dx|y)$ such that

$$P_X(dx) = \int K(dx|y) P_Y(dy), \qquad \int H_\mu(K(\cdot\,|y)) P_Y(dy) = E. \quad (2.6)$$

The first identity is marginal compatibility: mixing the conditional source distributions recovers the prescribed source distribution. The second identity fixes the average posterior entropy. On standard Borel spaces, the mixture defines a joint distribution and induces a forward test channel. Conversely, a source distribution and a forward test channel induce a posterior that automatically satisfies marginal compatibility.

**Theorem 1 (Attainability and explicit form of the rate–entropy function).**

Fix the source distribution and allow the representation alphabet and randomized test channel to vary. If $X$ is discrete with finite entropy, then every $E \in [0, H(X)]$ is attainable, and

$$R_e(E) = H(X) - E, \quad 0 \le E \le H(X); \qquad R_e(E) = 0, \quad E \ge H(X). \ (2.7)$$

If $X \in \mathbb{R}^m$ is absolutely continuous and $\int p_X(x)|\log p_X(x)|\,dx < \infty$, then every finite $E < h(X)$ is attainable, and

$$R_e(E) = h(X) - E, \quad -\infty < E < h(X); \qquad R_e(E) = 0, \quad E \ge h(X). \ (2.8)$$

Proof. For a discrete source, time-share between the identity channel and complete erasure. Independently of $X$, reveal $X$ with probability $\lambda$ and output an erasure symbol outside the source alphabet with probability $1 - \lambda$. Since the two events are distinguishable at the output, $H(X|Y) = (1 - \lambda)H(X)$ and $I(X;Y) = \lambda H(X)$. Varying $\lambda \in [0,1]$ gives the entire interval. A constant output achieves zero rate for budgets at least $H(X)$. This agrees with the rate–distortion theory under logarithmic loss [16]–[18].

For an absolutely continuous source, use time sharing between a partition channel and complete erasure. Absolute continuity makes the source measure nonatomic. Thus, for any integer $M \ge 2$, there is a measurable partition $A_1, \dots, A_M$ with $P_X(A_j) = 1/M$. Let $J$ be the cell index, so that $H(J) = \log M$. Independently of $X$, output $J$ with probability $\lambda$ and the erasure symbol $\perp$ otherwise. Then

$$I(X;Y) = \lambda \log M, \qquad h(X|Y) = h(X) - \lambda \log M. \ (2.9)$$

The output probabilities are $P(Y = \perp) = 1 - \lambda$ and $P(Y = j) = \lambda/M$. The posterior is $p_X$ on erasure and $p_{X|Y=j}(x) = M p_X(x) \mathbf{1}_{A_j}(x)$ on observing cell index $j$, whenever that output has positive probability. Their mixture recovers $p_X$. Log-density integrability ensures that the entropies in (2.9) are finite.

Given a finite $E \le h(X)$, choose an integer $M \ge 2$ with $\log M \ge h(X) - E$ and set $\lambda = [h(X) - E]/\log M$. This gives $h(X|Y) = E$, and (2.5) yields (2.8). For $E \ge h(X)$, a constant output is feasible and achieves zero rate. This completes the proof.

Both constructions use time sharing. In the continuous case, the number of partition cells determines the attainable entropy interval, and the sharing probability fills that interval. No second-moment assumption is required, so the result includes Cauchy sources. The lossless endpoint for a discrete source is $E = 0$; for a continuous source, $R_e(E) \to \infty$ as $E \to -\infty$. On the attainable interval below the source entropy, the rate–entropy function is a line of slope $-1$.

## 2.5 From Physical Distortion Budgets to Entropy Budgets

Conditional entropy provides a common information-based distortion coordinate. We use physical distortion to refer to a task-dependent cost, as distinct from entropy distortion. For a fixed source–representation distortion function $d$, define the maximum conditional entropy permitted by budget $D$ as

$$E_d(D) = \sup_{P_{Y|X}:\ \mathbb{E}d(X,Y) \le D} H_\mu(X|Y). \ (2.10)$$

On an interval where the feasible set is nonempty, the entropy-difference representation of mutual information is valid, and the relevant optimal values are finite,

$$R_d(D) = H_\mu(X) - E_d(D) = R_e(E_d(D)). \ (2.11)$$

The first equality follows from the mutual information identity. The second follows from Theorem 1 when $E_d(D)$ belongs to its applicable entropy range. These are identities between optimal values and do not require the supremum defining $E_d(D)$ to be attained. They separate the calculation into $D \mapsto E_d(D)$ and $E \mapsto R_e(E)$. The former depends on the source and the task cost; the latter has a common linear form. Section 3 develops conditions and constructions for attaining the maximum conditional entropy.

The function $E_d(D)$ is nondecreasing and concave wherever it is finite. Monotonicity follows from inclusion of the feasible sets. Concavity follows from concavity of conditional entropy in the joint distribution with the source marginal fixed, or equivalently from convexity of the rate–distortion function. At differentiability points, $R_d'(D) = -E_d'(D)$. Figure 1 compares two source–distortion pairs.

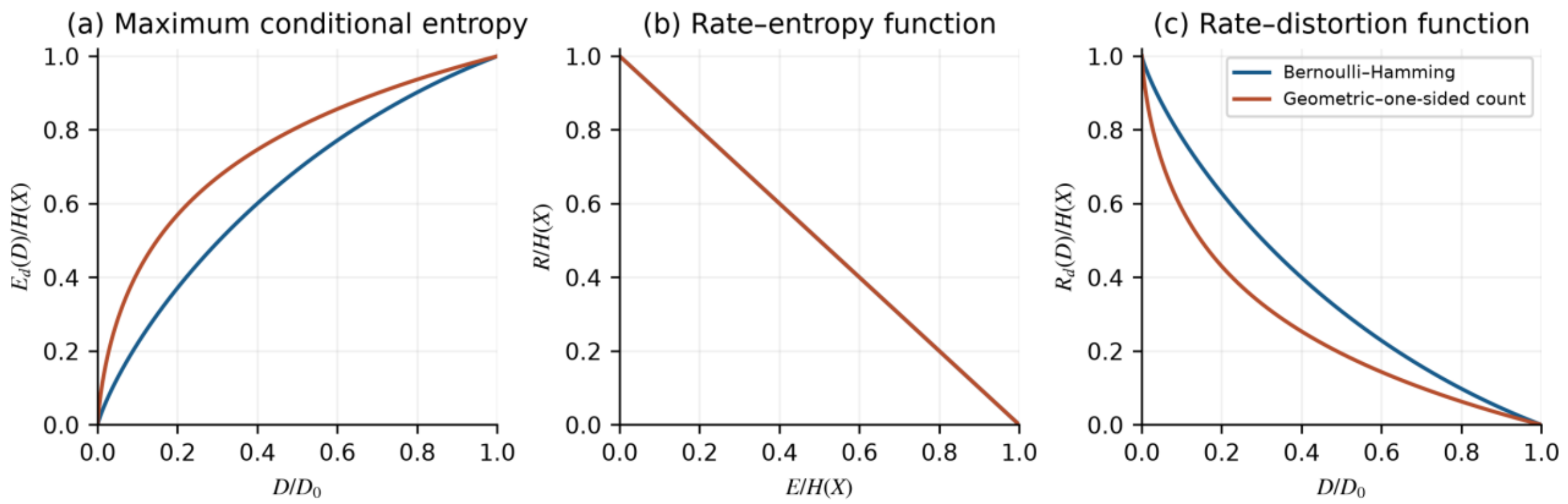

Fig. 1. Physical distortion budgets mapped to a common entropy coordinate. The Bernoulli source has parameter $p = 0.3$ and Hamming distortion, with zero-rate budget $D_0 = 0.3$. The geometric source has mean 10 and one-sided counting distortion, with $D_0 = 10$. Budgets and rates are normalized by the corresponding zero-rate budget and source entropy. Distinct functions $E_d(D)$ map the two rate–distortion curves to the same rate–entropy line.

## 3 Conditions for Constructing Closed-Form Rate–Distortion Functions

### 3.1 Posterior Representation of Test Channels

Fix a source distribution $P_X \ll \mu$ and write $p_X = dP_X/d\mu$. The source and representation alphabets are standard Borel spaces, and $\mu$ is a $\sigma$-finite reference measure. For the channels considered here, assume that the posterior distributions admit densities with respect to the reference measure. Such a channel has a posterior representation $(P_Y, K)$, where $P_Y$ is its induced output distribution and $K(\cdot\,|y)$ is the posterior density with respect to $\mu$. They satisfy

$$\int K(x|y)\mu(dx) = 1 \quad (P_Y \text{ a.e.}), \; (3.1)$$

$$p_X(x) = \int K(x|y)P_Y(dy) \quad (\mu \text{ a.e.}). \; (3.2)$$

Equation (3.2) expresses marginal compatibility with the prescribed source. A pair satisfying (3.1)–(3.2) defines the joint distribution $P_{XY}(dx, dy) = K(x|y)\mu(dx)P_Y(dy)$ and, wherever $p_X(x) > 0$, the forward test channel

$$P_{Y|X}(dy|x) = \frac{K(x|y)}{p_X(x)} P_Y(dy). \; (3.3)$$

All test channels considered below satisfy (3.1)–(3.2). Marginal compatibility is a prerequisite for a valid channel; optimality is assessed among channels with this prescribed source marginal.

All entropy differences below are understood as differences of finite quantities, with the required log densities integrable. Finite mutual information implies absolute continuity of the posterior with respect to the source law almost surely. Channels with infinite mutual information cannot improve a finite candidate optimum.

Fix a measurable distortion function $d: \mathcal{X} \times \mathcal{Y} \to [0, +\infty]$. We use $E_d(D)$ and $R_d(D)$ as defined in Section 2 and retain the integrability assumptions stated there for source entropy, conditional entropy, and the relevant expectations. This section establishes optimality and parametric representations for a fixed distortion function. Continuous logarithmic loss, considered in Section 3.3, may be negative and is used when the relevant expectations are finite.

The classical Shannon–Berger parametric formulation describes a forward channel using a reference output distribution $Q$. At an optimum, $Q = P_Y$, and Bayes' rule gives the posterior representation [2], [3], [5]. Here the unknowns are directly $(P_Y, K)$; the classical test function corresponds to the integral of the Gibbs kernel over the source variable.

### 3.2 Conditions for Attaining Maximum Conditional Entropy

**Theorem 2 (Attainment of maximum conditional entropy).**

Let $(P_Y, K)$ be a marginally compatible test channel with finite average distortion $D$. If there exist $\beta > 0$ and a measurable function $A_\beta$ with $\mathbb{E}|A_\beta(X)| < \infty$ such that the Gibbs kernel

$$K_\beta(x|y) = \exp\{-\beta d(x,y) - A_\beta(x)\} \quad (3.4)$$

satisfies

$$K(x|y) = K_\beta(x|y) \quad (\mu \otimes P_Y \text{ a.e.}), \quad (3.5)$$

$$a_\beta(y) := \int K_\beta(x|y)\mu(dx) \leq 1, \qquad y \in \mathcal{Y}, \quad (3.6)$$

then this channel attains the maximum conditional entropy and the rate–distortion function at budget D, with

$$H_\mu(X|Y) = E_d(D) = \beta D + \mathbb{E}A_\beta(X), \quad (3.7)$$

$$R_d(D) = H_\mu(X) - E_d(D), \qquad -\beta \in \partial R_d(D). \quad (3.8)$$

Here $-\beta \in \partial R_d(D)$ means that $R_d(D') \geq R_d(D) - \beta(D' - D)$ for every feasible budget $D'$.

Proof. Consider any marginally compatible channel with output $Y'$, posterior density $r(x|y)$, and actual distortion $D_{\text{act}}$. Jensen's inequality and (3.6) imply, for almost every output,

$$\int r(x|y)\log\frac{K_\beta(x|y)}{r(x|y)}\,\mu(dx) \leq \log\int_{\{x:r(x|y)>0\}} K_\beta\,(x|y)\,\mu(dx) \leq 0. \quad (3.9)$$

Expanding the left-hand side and averaging over $Y'$ gives the following bound. Since the source marginal is fixed, $\mathbb{E}A_\beta(X)$ is the same for every competing channel:

$$H_\mu\big(X|Y^{'}\big) \leq \beta D_{act} + \mathbb{E}A_\beta(X). \quad (3.10)$$

If $D_{\text{act}} \leq D$, the right-hand side is at most $\beta D + \mathbb{E}A_\beta(X)$. For the candidate channel, substituting (3.5) into the definition of conditional entropy gives $H_\mu(X|Y) = \beta D + \mathbb{E}A_\beta(X)$. Thus the channel attains the maximum conditional entropy, proving (3.7); the mutual information identity gives the corresponding rate. For any feasible budget $D'$, (3.10) also yields

$$R_d\big(D^{'}\big) \geq H_\mu(X) - \beta D^{'} - \mathbb{E}A_\beta(X) = R_d(D) - \beta\big(D^{'} - D\big).$$

Hence $-\beta$ is a supporting slope at this point, completing the proof of (3.8).

The exponential form agrees with the Gibbs distributions arising in the maximum-entropy principle [19]. For a prescribed source, it must additionally satisfy marginal compatibility and the kernel mass bound for every output.

At the operating point of Theorem 2, any compatible channel $(P_{Y'}, r)$ with $D_{\text{act}} \leq D$ has a rate gap equal to its conditional entropy gap:

$$I(X;Y') - R_d(D) = E_d(D) - H_\mu(X|Y'). \quad (3.11)$$

Expanding the posterior log ratio using (3.4) and substituting (3.7) gives

$$E_d(D) - H_\mu\big(X|Y^{'}\big) = \beta(D - D_{act}) + \mathbb{E}_{Y^{'}}\int r\big(x|Y^{'}\big)\log\frac{r(x|Y^{'})}{K_\beta(x|Y^{'})}\,\mu(dx). \quad (3.12)$$

The two terms are the distortion-budget slack and the posterior mismatch. Both are nonnegative, by $D_{\text{act}} \leq D$ and (3.9), respectively. The mismatch vanishes exactly when $r(\cdot\,|Y') = K_\beta(\cdot\,|Y')$ holds $\mu$-almost everywhere for almost every actual output; the kernel then has unit mass at those outputs. Consequently, the rate gap is zero exactly when the budget is saturated and the posteriors match. If the kernel is normalized at the actual outputs, the mismatch is an average Kullback–Leibler divergence. Otherwise, it is the log-ratio integral in (3.12). Even when no compatible channel attaining the bound has been found, the identity remains valid with $E_d(D)$ on the left replaced by $\beta D + \mathbb{E}A_\beta(X)$.

For a fixed distortion function, the candidate must satisfy the source marginal equation, the Gibbs form, and the mass bound over the entire output alphabet. Once (3.6) holds, the remaining task is to find a probability distribution $P_Y$, supported on outputs with $a_\beta(y) = 1$, that solves (3.2). Section 4 realizes this mixture equation through convolution decompositions. Section 5 starts from a forward test channel and the source distribution, obtains a compatible posterior, and then checks optimality. Section 4.1 further decomposes posterior mismatch into residual dependence and residual distribution mismatch.

### 3.3 Parametric Representation of Rate–Distortion Functions

Fix $P_X$, $\mathcal{Y}$, and $d$. If a compatible family $\{(P_{Y,t}, K_t): t \in T\}$ satisfies Theorem 2 at every parameter value, with multiplier $\beta(t)$, then the quantities

$$D(t) = \mathbb{E}_t d(X,Y), \qquad E(t) = H_\mu(X|Y_t), \quad (3.13)$$

determine the rate–distortion function by

$$R_d(D(t)) = H_\mu(X) - E(t), \qquad t \in T. \quad (3.14)$$

Equations (3.13)–(3.14) describe the curve over $D(T)$. If $D(t)$ is invertible, eliminating $t$ yields an explicit expression for $R_d(D)$; otherwise, the parametric expression is retained with its parameter domain and distortion range. An expression involving finitely many operations and known special functions provides a closed-form parametrization, whereas an explicit convergent series provides a computable parametrization. Endpoints are obtained from feasible channels or as limits.

For multiple constraints, fix $d_1, \ldots, d_m$ and replace $\beta d$ in (3.4) by $\sum_j \beta_j d_j$. Upper-bound constraints $\mathbb{E}d_j \leq D_j$ require $\beta_j \geq 0$ and complementary slackness $\beta_j(D_j - \mathbb{E}d_j) = 0$ for the candidate channel. If posterior matching and the global kernel mass bound hold, then

$$R_{\boldsymbol{d}}(\boldsymbol{D}) = H_\mu(X) - \sum_{j=1}^{m} \beta_j D_j - \mathbb{E}A(X). \quad (3.15)$$

The proof follows the same Jensen argument. Replacing $\beta d$ by $\sum_j \beta_j d_j$ gives $H_\mu(X|Y') \leq \sum_j \beta_j D_{j,\mathrm{act}} + \mathbb{E}A(X)$. The budget constraints and nonnegative multipliers bound this by $\sum_j \beta_j D_j + \mathbb{E}A(X)$. Posterior matching and complementary slackness give equality for the candidate, proving (3.15). Multipliers for equality constraints may be arbitrary real numbers, since the corresponding expectations are fixed. Varying the budgets for a fixed collection of costs gives a rate–distortion surface; specifying relations among the budgets selects a slice of that surface.

Under logarithmic loss, a representation $y(\cdot)$ is itself a normalized density with respect to $\mu$, and $d_{\log}(x,y) = -\log y(x)$. Taking $\beta = 1$ and $A \equiv 0$ gives $K(x|y) = y(x)$ in (3.4), and (3.6) holds with equality for every representation. Using the compatible posteriors constructed in Section 2 as density-valued representations gives, throughout the corresponding attainable entropy range,

$$D_{\log} = H_\mu(X|Y) = E, \qquad R_{\log}(E) = R_e(E) = H_\mu(X) - E. \quad (3.16)$$

Theorem 1 establishes entropy attainability, while Theorem 2 with $\beta = 1$ and $A \equiv 0$ certifies optimality under logarithmic loss [4], [16]–[18]. If a strictly increasing function defines an alternative entropy-budget coordinate $\Delta = \varphi\left(H_\mu(X|Y)\right)$, then $R_\Delta(D) = R_e(\varphi^{-1}(D))$ on the corresponding range. Section 4.2.3 and the appendix give examples of such coordinates.

Such a scale budget is a functional of the joint distribution induced by the test channel. Its rate function is the infimum of mutual information over channels satisfying that functional constraint, whose feasible set is determined by conditional entropy. In contrast, (2.2) constrains the expectation of a fixed single-letter cost. The applications below specify whether a result concerns an expected cost or a transformed entropy budget.

### 3.4 Identifying a Fixed Distortion Function from Posterior Structure

Fix the source, representation alphabet, and reference measure, and consider a compatible posterior family with a common support. If the full negative log density at every parameter value decomposes into a positive multiplier times the same nonnegative source–representation function, plus an integrable term depending only on the source variable, then that common function is a candidate fixed distortion. Writing this decomposition in the form (3.4) and verifying (3.6) for every output permits application of Theorem 2 at each parameter value. Equations (3.13)–(3.14) then determine the distortion budget and rate. The kernel outside the actual output support must also be defined using the same distortion function.

The identification uses the parameter dependence of the posterior family. As the parameter varies, the distortion function remains fixed, while the multiplier and source-only term may change. A common set of forbidden source–representation pairs may be assigned infinite cost. The negative log posterior at a single parameter value supplies a candidate cost only at that point. A full curve requires the fixed-cost representation, the kernel mass bound, and the range of marginal compatibility. Positive rescaling of a candidate cost changes only the budget units; physical interpretation and task requirements also inform the final choice of distortion.

Along the independent-residual path, the negative log posterior reduces to the negative log residual density, with quadratic, absolute-value, linear, and logarithmic terms yielding different error costs. Along the conjugate-posterior path, terms coupling the source and representation variables reveal costs such as relative error. Conditional entropy determines the rate, while the full posterior structure identifies the distortion. Sections 4 and 5 carry out these steps and verify optimality under the specified budgets.

# 4 Rate–Distortion Functions for Generalized Additive Sources

This section constructs compatible posteriors through independent decompositions, identifies distortion functions from the negative log residual density, and verifies optimality. Additive decompositions are treated first, followed by multiplicative decompositions in logarithmic coordinates. We write $Y$ for the representation, $\hat{X} = g(Y)$ for the point reconstruction, and $Z$ for the residual. Logarithms are natural and rates are in nats.

## 4.1 Convolution Decompositions and Rate–Distortion Construction

**Definition 4.1 (Generalized additive source).**

Here the operation $\circ$ is addition or multiplication, as specified below.

Fix a distribution family $\mathcal{P}$ and a source distribution $P_X \in \mathcal{P}$. If there exist independent, nondegenerate random variables $\hat{X}$ and $Z$, with $P_Z \in \mathcal{P}$, such that $\hat{X} \circ Z$ has distribution $P_X$, then the source admits a generalized additive decomposition. We consider addition and multiplication. The additive decomposition is

$$X \overset{d}{=} \hat{X} + Z, \qquad \hat{X} \perp Z. \quad (4.1)$$

For positive random variables, the multiplicative decomposition is

$$X \overset{d}{=} \hat{X} Z, \qquad \hat{X} \perp Z. \quad (4.2)$$

The admissible parameter set is determined by independence, membership of the residual distribution in $\mathcal{P}$, and the distributional identity. Its range is specified in each example. Decompositions are nontrivial at interior parameter values; degenerate components are allowed at endpoints. The reconstruction distribution need not belong to $\mathcal{P}$.

Gaussian, Cauchy, Laplace, uniform, Gamma, negative binomial, geometric, and Poisson sources admit additive decompositions; Pareto and zeta sources admit multiplicative decompositions. A residual in the same family describes the source decomposition. Optimality under an average-cost constraint additionally requires a maximum-entropy residual. The maximum-entropy argument below also applies when the source and residual belong to different families.

An additive decomposition is equivalent to the convolution identity $P_X = P_{\hat{X}} * P_Z$ [20]. With $\hat{X} = g(Y)$ and $Z \perp Y$, the backward kernel is $K(B|y) = P_Z\big(B - g(y)\big)$ for measurable $B$. Integrating over the induced output distribution gives $\int K(B|y)P_Y(dy) = P_X(B)$, so the decomposition ensures marginal compatibility. When the residual has a density or probability mass function, $K(x|y) = p_Z(x - g(y))$.

Given the source distribution and a candidate residual distribution, the product identity for characteristic functions determines the reconstruction distribution and the admissible decomposition range. Wherever the residual characteristic function is nonzero,

$$\varphi_{\hat{X}}(t) = \frac{\varphi_X(t)}{\varphi_Z(t)}. \quad (4.3)$$

The quotient must be continuous, positive definite, and equal to one at the origin to be a characteristic function. If the denominator has zeros, one instead solves $\varphi_X = \varphi_{\hat{X}}\varphi_Z$ directly. For nonnegative variables, a Laplace transform may be used. For nonnegative integer-valued variables, the quotient of probability generating functions must have nonnegative power-series coefficients summing to one.

Once a convolution decomposition supplies a compatible posterior, the conditions of Theorem 2 apply. Let the additive space be $\mathbb{R}^m$ or $\mathbb{Z}^m$, with Lebesgue or counting measure, respectively. A nonnegative residual cost $d(z)$ induces the source–representation distortion

$$d(x, y) = d(x - g(y)). \quad (4.4)$$

The one-argument and two-argument uses of $d$ denote the same cost in residual and source–representation coordinates. Forbidden residuals have infinite cost. Since the representation enters only through $g(Y)$, constructions using reconstruction values as representations may take $Y = \hat{X}$. The relevant entropies, mutual information, and expectations are assumed well defined and finite.

If the independent decomposition $X = g(Y_t) + Z_t$ has a residual family with common support satisfying

$$-\log p_{Z_t}(z) = \beta(t)d(z) + A(t), \qquad \beta(t) > 0, \quad (4.5)$$

where $d(z) \geq 0$ is independent of $t$ and is infinite outside the common support, then the compatible posterior $K_t(x|y) = p_{Z_t}(x - g(y))$ satisfies the kernel mass condition of Theorem 2 by translation invariance. It has unit mass over the full additive space and mass at most one when restricted to the source support. Hence, for the fixed distortion $d(x, y) = d(x - g(y))$, throughout the admissible decomposition range,

$$D(t) = \mathbb{E}d(Z_t), \quad E_d(D(t)) = H_\mu(Z_t), \quad R_d(D(t)) = H_\mu(X) - H_\mu(Z_t). \quad (4.6)$$

Equation (4.5) identifies a fixed residual cost, and (4.6) gives its exact rate–distortion parametrization. If $D(t)$ is invertible, the parameter can be eliminated.

Squared error corresponds to a Gaussian residual, absolute error to a Laplace residual, nonnegative linear error to an exponential residual, and counting error to a geometric residual. The same maximum-entropy argument certifies optimality in each case. The source-specific question is whether the required independent decomposition exists and over which parameter range.

**Rate gaps and entropy gaps.**

Fix a parameter value and write $D = D(t)$, $Z_D = Z_t$, and $\beta = \beta(t)$. For any marginally compatible channel with actual distortion $D_{\mathrm{act}} \leq D$, finite mutual information, and finite residual entropy, define $Z = X - g(Y)$. Since $H_\mu(X|Y) = H_\mu(Z|Y)$, the gap identity in Section 3 becomes

$$I(X;Y) - R_d(D) = I(Z;Y) + H_\mu(Z_D) - H_\mu(Z). \quad (4.7)$$

The residual dependence and residual entropy gap are both nonnegative. Substitution of the negative log Gibbs density further gives

$$H_\mu(Z_D) - H_\mu(Z) = D_{\mathrm{KL}}(P_Z \| P_{Z_D}) + \beta(D - D_{\mathrm{act}}). \quad (4.8)$$

Thus, the residual entropy gap consists of distribution mismatch and budget slack. The rate gap vanishes exactly when the residual is independent of the representation, its distribution is $P_{Z_D}$, and the budget is saturated. For an independent-residual channel, the rate gap equals the residual entropy gap.

**A common derivation of established results.**

Gaussian sources under quadratic distortion and Laplace sources under absolute error are classical examples [2], [3]; the Laplace decomposition is also discussed in [21]. The Hamming-distortion result for a uniform source on a finite group has an analogous independent-residual interpretation [2], [3]. For vector Gaussian sources, matrix budgets constrain directional error [2], [3], whereas determinant budgets constrain error volume [11]; these constraints define different feasible sets. The authors' earlier work [4] relates conditional entropy to physical scales for uniform, Cauchy, and vector Gaussian sources. Its entropy widths and entropy scales are reparametrizations of conditional entropy budgets, whereas the Cauchy strength budget is defined through the error distribution [22]. The one-sided logarithmic multiplicative distortion result for a Pareto source follows from Verdú's exponential-source result [6] by a logarithmic change of variables and is another instance of the present construction. The appendix table collects the formulas, budgets, and sources.

The determinant budget in [11] uses the dimension-root of the error covariance determinant; the appendix uses the determinant itself. Correcting a constant error bias preserves mutual information and reduces the error second-moment matrix to its covariance matrix. Consequently, the second-moment determinant constraint in the appendix has the same optimal value as the covariance-determinant constraint. An independent Gaussian residual with covariance proportional to the source covariance attains this value.

## 4.2 Examples of Additive Sources

Gamma sources under one-sided mean error and negative binomial sources under one-sided counting distortion illustrate the range over which a maximum-entropy residual can be realized. The geometric source is included as a special case of the negative binomial family. For point-reconstruction examples, take $Y = \hat{X}$.

### 4.2.1 Gamma Sources Under One-Sided Mean Error

Let $X \sim \mathrm{Gamma}(r, \theta)$, where $r \geq 1$ and $\theta > 0$ is the scale parameter. Its differential entropy is

$$h(X) = \kappa(r) + \log\theta, \qquad \kappa(r) = r + \log\Gamma(r) + (1 - r)\psi(r), \tag{4.9}$$

where $\psi$ is the digamma function. For every $0 < D \leq \theta$, there is an independent decomposition

$$X = \hat{X}_D + Z_D, \quad \hat{X}_D \perp Z_D, \tag{4.10}$$

where $Z_D$ is exponential with mean $D$.

The reconstruction has Laplace transform

$$\mathcal{L}_{\hat{X}_D}(u) = \frac{1+Du}{(1+\theta u)^r} = \left(1 - \frac{D}{\theta}\right)(1 + \theta u)^{-r} + \frac{D}{\theta}(1 + \theta u)^{-(r-1)}. \tag{4.11}$$

The right-hand side is a mixture of $\mathrm{Gamma}(r, \theta)$ and $\mathrm{Gamma}(r - 1, \theta)$ with weights $1 - D/\theta$ and $D/\theta$. For $r = 1$, the second component is a point mass at zero. This decomposition establishes marginal compatibility. Consider the one-sided error

$$d(x, \hat{x}) = \begin{cases} x - \hat{x}, & 0 \leq \hat{x} \leq x, \\ +\infty, & \text{otherwise.} \end{cases} \tag{4.12}$$

Among nonnegative residuals with mean at most D, the exponential distribution with mean D maximizes differential entropy:

$$\sup_{Z \geq 0,\, \mathbb{E}Z \leq D} h(Z) = h(Z_D) = 1 + \log D. \tag{4.13}$$

The decomposition realizes this maximum-entropy residual. Taking $Y = \hat{X}_D$, Theorem 2 and (4.6) give

$$R_d(D) = h(X) - 1 - \log D = \kappa(r) - 1 + \log\frac{\theta}{D}, \qquad 0 < D \leq \theta. \tag{4.14}$$

For $r = 1$, this expression covers the entire positive-rate region. For $r > 1$, it gives the exact rate–distortion function on the interval where the maximum-entropy lower bound is attained. The zero-rate threshold is $D = r\theta$; the interval $\theta < D < r\theta$ is not determined by this construction.

Setting $r = \nu/2$ and $\theta = 2$, with $\nu \geq 2$, gives the corresponding chi-square case. Section 5 keeps the Gamma source fixed and constructs posteriors from forward test channels, yielding results under Itakura–Saito distortion and entropy distortion for counting representations.

### 4.2.2 Negative Binomial Sources Under One-Sided Counting Distortion

Let $X \sim \mathrm{NB}(r, p)$, with $r > 0$ and $0 < p < 1$, using the convention $p_X(k) = \Gamma(k + r)p^r(1 - p)^k/[\Gamma(r)k!]$, $k = 0,1, \ldots$. Choose a geometric residual of mean $D > 0$ and consider

$$X = \hat{X}_D + Z_D, \qquad \hat{X}_D \perp Z_D, \qquad p_{Z_D}(k) = \frac{1}{1+D}\left(\frac{D}{1+D}\right)^k, \quad k \geq 0. \quad (4.15)$$

The geometric distribution is a negative binomial distribution of shape one, so the residual belongs to the negative binomial family. The quotient of probability generating functions yields the required reconstruction probabilities:

$$p_{\hat{X}_D}(k) = (1 + D)p_X(k) - Dp_X(k - 1), \qquad p_X(-1) = 0. \quad (4.16)$$

These coefficients sum to one. The ratio of adjacent source probabilities,

$$\frac{p_X(k)}{p_X(k-1)} = (1 - p)\frac{k+r-1}{k}, \qquad k \geq 1, \quad (4.17)$$

shows that all reconstruction probabilities are nonnegative if and only if

$$0 \leq D \leq D_*, \qquad D_* = \begin{cases} \frac{1-p}{p}, & r \geq 1, \\ \frac{r(1-p)}{1-r(1-p)}, & 0 < r < 1. \end{cases} \quad (4.18)$$

The independent decomposition therefore exists throughout this interval, with a zero residual at $D = 0$. For $D > 0$, the negative log residual probability is

$$-\log p_{Z_D}(k) = \log(1 + D) + k\log\frac{1+D}{D}. \quad (4.19)$$

This form identifies the one-sided counting distortion

$$d_{cnt}(x, \hat{x}) = \begin{cases} x - \hat{x}, & 0 \leq \hat{x} \leq x, \\ +\infty, & \text{otherwise.} \end{cases} \quad (4.20)$$

The geometric residual maximizes entropy among nonnegative integer-valued distributions of mean at most D, and the independent decomposition ensures marginal compatibility. Theorem 2 and (4.6) therefore yield

$$R_{cnt}(D) = H(X) - H_G(D), \qquad 0 \leq D \leq D_*, \quad (4.21)$$

where $H_G(D) = (D + 1)\log(D + 1) - D\log D$, with $0\log 0 = 0$. For $D \geq r(1 - p)/p$, the constant reconstruction zero achieves zero rate. If $r \neq 1$, the exact function on $D_* < D < r(1 - p)/p$ is not determined by this decomposition.

Equation (4.21) is exact on the stated interval. For general shape parameters, the source entropy is evaluated by the entropy series of the negative binomial probability mass function, which converges absolutely because of its exponential tail. Exactness here means that the mutual information optimization has been solved; the geometric case further gives an explicit expression with no source entropy left to evaluate.

For $r = 1$, the source is $\mathrm{Geom}(p)$ and $D_* = (1 - p)/p = \mathbb{E}X$. The result then covers the entire nonzero-rate range, giving

$$R_{cnt}(D) = \begin{cases} H_G((1 - p)/p) - H_G(D), & 0 \leq D \leq (1 - p)/p, \\ 0, & D \geq (1 - p)/p. \end{cases} \quad (4.22)$$

### 4.2.3 Poisson Sources Under a Residual-Count Entropy Budget

Let $X \sim \mathrm{Pois}(\lambda)$, with $\lambda > 0$, and write $H_P(t) = H(\mathrm{Pois}(t))$. This function is continuous and strictly increasing for $t \geq 0$, with $H_P(0) = 0$. Using the Poisson mean to express the remaining uncertainty, define the equivalent residual-count distortion functional

For $t > 0$, differentiation gives $H_P'(t) = \mathbb{E}\log(N_t + 1) - \log t$, where $N_t \sim \text{Pois}(t)$. Since $\mathbb{E}[1/(N_t + 1)] = (1 - e^{-t})/t$, Jensen's inequality yields $H_P'(t) \geq -\log(1 - e^{-t}) > 0$. Continuity at zero follows from $H_P(t) \to 0$ as $t \downarrow 0$.

$$D_P(X|Y) = H_P^{-1}\big(H(X|Y)\big). \quad (4.23)$$

This is the entropy-budget reparametrization of Section 3.3: mutual information is minimized over all test channels satisfying the budget defined by (4.23). The independent decomposition below gives this coordinate its mean residual-count interpretation.

The constraint $D_P(X|Y) \leq D$ is equivalent to $H(X|Y) \leq H_P(D)$. For every $0 \leq D \leq \lambda$, take

$$X = Y + Z_D, \quad Y \sim \text{Pois}(\lambda - D), \quad Z_D \sim \text{Pois}(D), \quad Y \perp Z_D. \quad (4.24)$$

Poisson additivity ensures marginal compatibility. Given $Y = y$, the posterior is the distribution of $y + Z_D$. Independence and translation invariance of discrete entropy give

$$H(X|Y) = H(Z_D) = H_P(D). \quad (4.25)$$

Thus, the decomposition attains the largest conditional entropy permitted by the budget. The mutual information identity yields the corresponding rate function:

$$R_P(D) = \begin{cases} H_P(\lambda) - H_P(D), & 0 \leq D \leq \lambda, \\ 0, & D \geq \lambda. \end{cases} \quad (4.26)$$

In this optimal decomposition, $D = \mathbb{E}Z_D$ is the mean residual count, linking the entropy budget to the counting parameter. At $D = 0$, the residual is zero and the rate is $H_P(\lambda)$. At $D = \lambda$, the representation is zero and the rate vanishes. A constant output remains feasible for larger budgets.

## 4.3 Log-Additive Sources and the Zeta Example

For a positive source, an independent multiplicative decomposition $X = \hat{X}Z$ becomes $\log X = \log\hat{X} + \log Z$. Marginal compatibility can therefore again be solved through convolution. In the original coordinates, the backward kernel is $K(B|y) = P_Z\big(B/g(y)\big)$, with $g(y) > 0$.

For continuous variables, a logarithmic transformation requires transforming both the density and the reference measure. A one-to-one transformation of a discrete variable preserves Shannon entropy. Optimality may be analyzed in the transformed coordinates or by directly checking the backward channel in the original coordinates.

**Zeta sources under divisibility-constrained multiplicative distortion.**

Let the source take values in the positive integers with probability mass function $p_X(n) = n^{-s}/\zeta(s)$, $s > 1$ [23]. Reconstructions are also positive integers. The negative log probability of a zeta residual consists of a log-residual term and a normalizing term. The multiplicative decomposition requires the reconstruction to divide the source value, giving the fixed distortion

$$d_\times(n, \hat{n}) = \begin{cases} \log(n/\hat{n}), & \hat{n} \mid n, \\ +\infty, & \text{otherwise.} \end{cases} \quad (4.27)$$

Let $Z_t \sim \text{zeta}(t)$, $t \geq s$. The Euler product $\zeta(s) = \prod_\ell (1 - \ell^{-s})^{-1}$, over primes $\ell$, shows that the prime exponents of the source are independent geometric random variables. Applying the geometric decomposition to each exponent gives an independent reconstruction with probabilities

$$p_{\hat{X}}(n) = \frac{\zeta(t)}{\zeta(s)} n^{-s} \prod_{\ell | n}(1 - \ell^{s-t}). \quad (4.28)$$

These probabilities are nonnegative. Since $\sum_\ell \ell^{-s} < \infty$, only finitely many prime factors in the random product are nontrivial almost surely, so it defines a positive integer-valued variable. The Euler product gives normalization and the mixture identity $p_X(n) = \sum_{a|n} p_{\hat{X}}(a) p_{Z_t}(n/a)$.

The backward channel is $K_t(n|\hat{n}) = \exp\{-t d_\times(n, \hat{n})\}/\zeta(t)$, normalized for every reconstruction value. Theorem 2 gives

$$D(t) = -\frac{\zeta'(t)}{\zeta(t)}, \qquad E(t) = \log\zeta(t) - t\frac{\zeta'(t)}{\zeta(t)}, \quad (4.29)$$

$$R_\times(D(t)) = H(X) - E(t), \qquad H(X) = \log\zeta(s) - s\frac{\zeta'(s)}{\zeta(s)}. \quad (4.30)$$

Since $D'(t) = -\mathrm{Var}(\log Z_t) < 0$, the parametrization covers $0 < D \le -\zeta'(s)/\zeta(s)$. At $t = s$, the reconstruction is identically one. As $t \to \infty$, distortion tends to zero and the rate tends to $H(X)$. For larger budgets the rate is zero. The Euler product thus reduces marginal compatibility to geometric convolutions of the prime exponents.

The residual log probability, Euler product, and maximum conditional entropy condition determine the distortion, compatible output distribution, and optimal curve, respectively. Divisibility preserves the multiplicative structure of the positive integers, while the log quotient measures the unreconstructed multiplicative factor. This construction extends the analytic approach to discrete sources with a prime-factor decomposition.

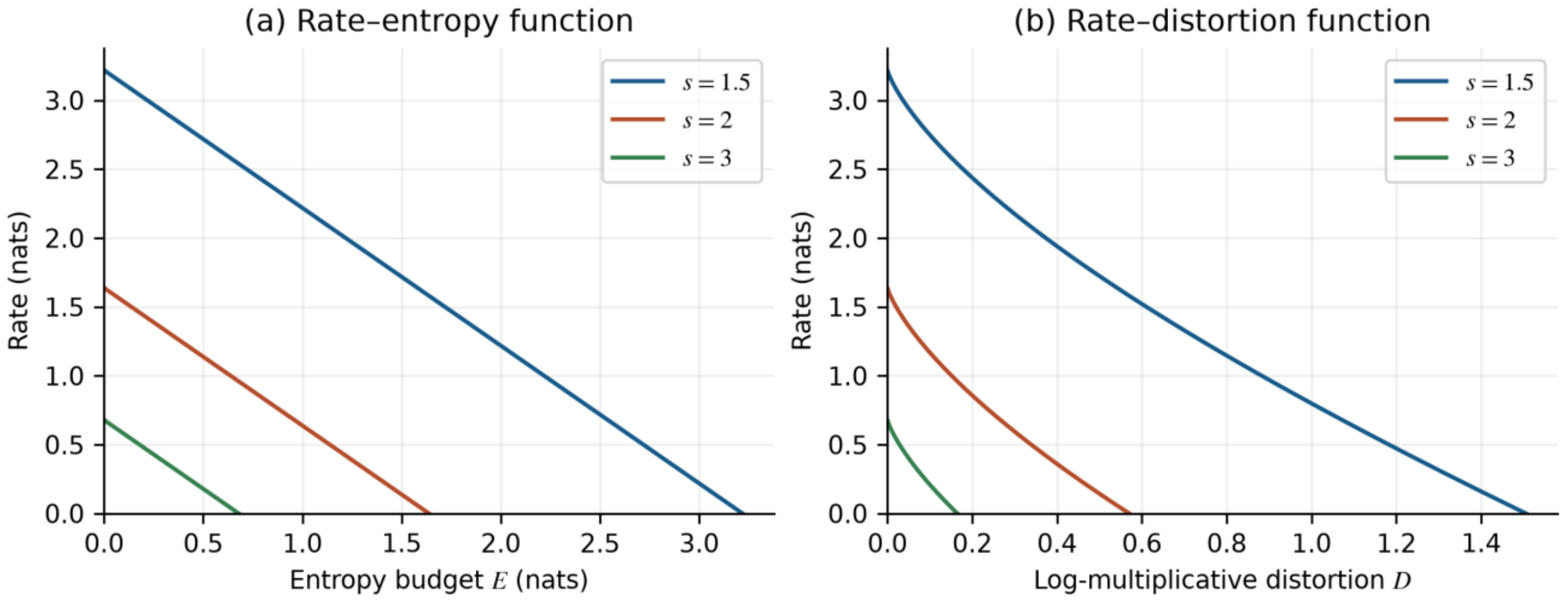

Fig. 2. Rate–entropy and rate–distortion functions for zeta sources with $s = 1.5, 2, 3$. Each curve fixes the source parameter $s$ and varies the residual parameter $t \ge s$. The horizontal coordinates are the entropy budget $E$ (left) and $D = \mathbb{E}\log(X/\hat{X})$ (right).

Both additive and log-additive decompositions construct compatible posteriors through independent residuals and determine the rate under the specified distortion constraint. The next section develops conjugate posteriors from given forward test channels and studies representations in alphabets distinct from the source alphabet.

# 5 Conjugate Posteriors and Distinct Source and Representation Alphabets

## 5.1 Conjugate Test Channels and Representation Alphabets

Fix the source distribution $P_X$ and a forward test channel $P_{Y|X,t}$. The joint distribution determines the induced output distribution $P_{Y,t}$ and posterior $K_t$, which satisfy

$$P_X(dx)P_{Y|X,t}(dy|x) = K_t(dx|y)P_{Y,t}(dy), \qquad P_X(dx) = \int K_t(dx|y)P_{Y,t}(dy). \quad (5.1)$$

Conjugacy means that the prior and posterior belong to the same distribution family; its general theory is developed in [24]. We consider Gamma and Dirichlet sources. Gamma sources are represented through reciprocal encoding with multiplicative noise and through counting channels. Dirichlet sources use multinomial counting, with Beta sources as the two-dimensional case. The established conjugate updates [25], [26] determine the output and posterior distributions from the source and forward channel. Their joint distribution ensures marginal compatibility.

Allowing distinct source and representation alphabets permits feature representations in a domain different from that of the source. Randomness in the encoder may produce different output features for the same input; multiplicative noise and random counts are two such mechanisms. The test channel specifies the input–output probability law, the posterior describes the remaining uncertainty about the source, and mutual information quantifies the source information retained by the representation.

For a compatible posterior, write $E(t) = H_\mu(X|Y)$. The rate–entropy relation of Section 2 gives

$$R_e\big(E(t)\big) = I(X;Y) = H_\mu(X) - E(t). \quad (5.2)$$

Each example specifies the test channel, induced output distribution, and posterior. The negative log posterior then identifies an appropriate fixed distortion, or the posterior entropy gives a rate–entropy parametrization. These test channels describe single-letter probabilistic representations. Section 6 provides block codes achieving the minimum rate under the same entropy budget. Operational achievability for a specified physical distortion follows the classical rate–distortion coding argument.

## 5.2 Gamma Sources Under Reciprocal Encoding with Multiplicative Noise

Let $X \sim \text{Gamma}(r,\theta)$, with $r,\theta > 0$. Here $X$ may represent an observable power quantity, while $r$ and $\theta$ are parameters of its statistical model. For example, the sum of squares of $\nu$ independent $\mathcal{N}(0,\sigma^2)$ amplitude components has distribution $\text{Gamma}(\nu/2, 2\sigma^2)$ and mean $\nu\sigma^2$. Normalization by $\sigma^2$ makes its mean equal to the number of degrees of freedom.

For $k > r$, let $a = k - r$. The encoder produces the reciprocal representation

$$Y = \frac{W}{X}, \qquad W \perp X, \qquad W \sim \text{Gamma}\left(a, \frac{1}{a}\right). \quad (5.3)$$

The random gain satisfies $\mathbb{E}W = 1$ and $\text{Var}(W) = 1/a$, so $\mathbb{E}[Y|X=x] = 1/x$. Thus, the output is a conditionally unbiased feature of reciprocal power, and $a$ controls the multiplicative noise level. Its conditional distribution is $Y|X = x \sim \text{Gamma}\big(a, 1/(ax)\big)$.

The Gamma conjugate update [25, Sec. 2.8.1] gives

$$X|Y = y \sim \text{Gamma}\left(k, \frac{\theta}{1+a\theta y}\right). \quad (5.4)$$

The induced output satisfies $a\theta Y \sim \text{BetaPrime}(k-r, r)$. The source is a power variable, whereas the representation is a continuous feature with a scaled beta-prime distribution. Given $y$, the posterior shape is fixed at $k$, and its inverse scale is updated from $\theta^{-1}$ to $\theta^{-1} + ay$. The feature thus determines the posterior scale of the power and the resulting point estimates.

Take the posterior mean $\hat{X} = k\theta/(1 + a\theta Y)$. For fixed $k > r$, this is a one-to-one function of $Y$, and the posterior becomes $K_k(\cdot\,|\hat{x}) = \text{Gamma}(k, \hat{x}/k)$. Expanding its negative logarithm gives

$$-\log K_k(x|\hat{x}) = k\left(\frac{x}{\hat{x}} - \log\frac{x}{\hat{x}} - 1\right) + \log x + k + \log\Gamma(k) - k\log k. \quad (5.5)$$

The term in parentheses identifies the fixed Itakura–Saito relative-error distortion

$$d_{IS}(x,\hat{x}) = \frac{x}{\hat{x}} - \log\frac{x}{\hat{x}} - 1, \qquad x,\hat{x} > 0. \quad (5.6)$$

We evaluate this distortion using reconstruction values as representations on the fixed alphabet $(0,\infty)$. The posterior-mean map converts the reciprocal feature into this representation and preserves mutual information at every nondegenerate parameter value. Varying the parameter changes the test channel while leaving the cost in (5.6) fixed.

The distortion is invariant under a common positive scaling of the source and reconstruction. The posterior kernel is normalized for every reconstruction in $(0,\infty)$ and satisfies Theorem 2 with multiplier $\beta = k$. Hence the compatible channel attains the IS rate–distortion function [13, Prop. 3].

Write

$$J(u) = u + \log\Gamma(u) - u\psi(u), \quad \kappa(u) = J(u) + \psi(u), \quad h(X) = \kappa(r) + \log\theta. \quad (5.7)$$

where $\psi = \Gamma'/\Gamma$. Since $X/\hat{X}|Y \sim \text{Gamma}(k, 1/k)$, the tower property for the posterior log moment gives

$$D(k) = \log k - \psi(k), \qquad E(k) = J(k) + \psi(r) + \log\theta, \quad (5.8)$$

$$R_{IS}\big(D(k)\big) = J(r) - J(k), \qquad k \geq r. \quad (5.9)$$

At $k = r$, take $Y = 0$ and $\hat{X} = r\theta$; the rate is zero for $D \geq D(r)$. Since $\psi'(k) > 1/k$, the function $D(k)$ decreases strictly to zero, covering the full positive-rate region. Moreover, $J'(k) = kD'(k)$, so the curve has slope $-k$. The scale term $\log\theta$ cancels between the prior and posterior entropies, and the rate–distortion function depends only on the shape $r$.

The conditional distribution of $Z = X/\hat{X}$ does not depend on $Y$, so $Z \perp \hat{X}$. The resulting independent product decomposition $X = \hat{X}Z$ connects this example to the log-additive construction of Section 4.

This derivation starts from a compatible conjugate posterior, identifies the fixed IS cost in its full negative log density, and verifies optimality through kernel normalization. Buzo et al. obtained the Gamma–IS result by solving the multiplicative mixture equation with Mellin transforms [13, Prop. 3 and Appendix]. Here the conjugate posterior and independent multiplicative decomposition describe the same optimal joint distribution, connecting the two construction paths.

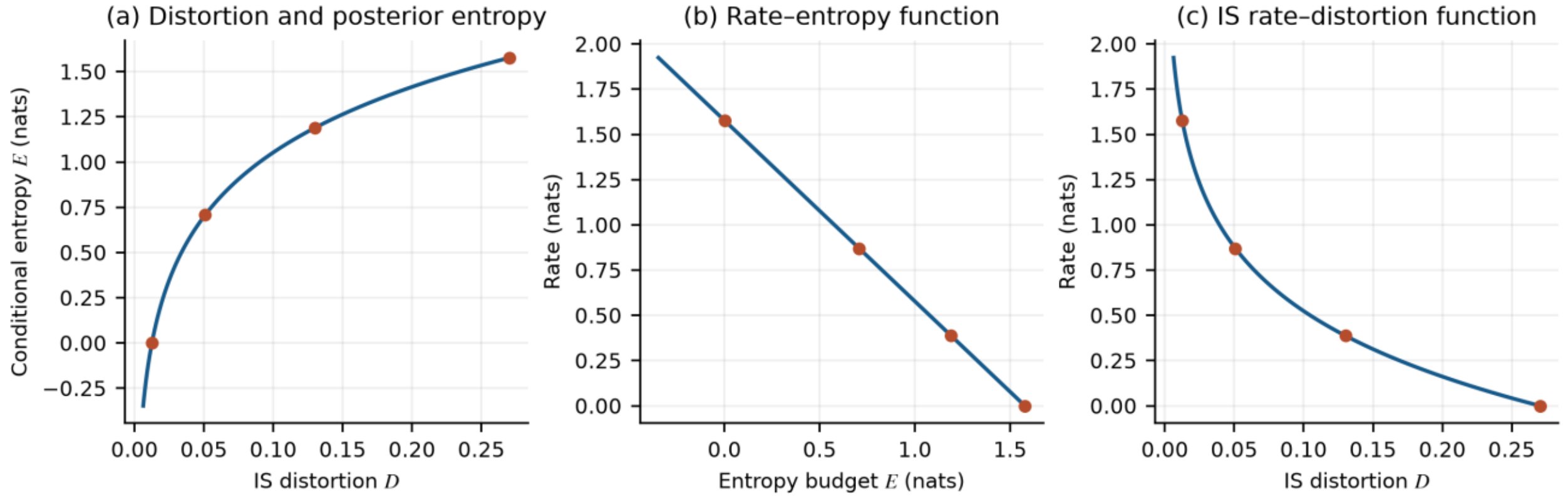

Fig. 3. Gamma-source representations in IS-distortion and entropy-distortion coordinates, with $r = 2$ and $\theta = 1$. Markers correspond to $k = 2,4,10,40$. All three panels use the same family of optimal test channels.

## 5.3 Gamma Sources Under Counting Encoding

Let $X \sim \mathrm{Gamma}(r, \theta)$, with $r, \theta > 0$, and let $N(\cdot)$ be a unit-rate Poisson process independent of $X$. For $t > 0$, the encoder outputs

$$Y = \frac{N(tX)}{t}, \qquad \mathbb{E}[Y|X] = X, \qquad \mathrm{Var}(Y|X) = \frac{X}{t}. \quad (5.10)$$

The output is a conditionally unbiased counting representation with the same units as the source, supported on a nonnegative grid of spacing $1/t$. The parameter $t$ controls counting precision. The Gamma–Poisson update [25, Sec. 2.2] gives the induced output distribution

$$tY \sim \mathrm{NB}\left(r, \frac{1}{1+\theta t}\right), \quad (5.11)$$

and posterior

$$X|Y = y \sim \mathrm{Gamma}\left(r + ty, \frac{\theta}{1+\theta t}\right). \quad (5.12)$$

Using $\kappa$ from Section 5.2 and averaging over the actual output gives

$$E(t) = \mathbb{E}_Y \kappa(r + tY) + \log\frac{\theta}{1+\theta t}, \quad (5.13)$$

$$R_e\big(E(t)\big) = \kappa(r) + \log\theta - E(t). \quad (5.14)$$

At $t = 0$, define the output to be identically zero, yielding $E(0) = h(X)$ and zero rate. For each fixed $t > 0$, the normalized and raw counts are in one-to-one correspondence and retain the same mutual information. The Gaussian entropy bound, followed by averaging the posterior variance, gives

$$E(t) \le \frac{1}{2}\log\frac{2\pi e r\theta^2}{1+\theta t} \to -\infty. \quad (5.15)$$

The parametrization therefore covers every finite $E \le h(X)$.

Continuity and strict monotonicity follow directly from the counting model. On each bounded parameter interval, the negative binomial output has a uniformly exponential tail, whereas the posterior entropy grows at most logarithmically with the count. The series for average posterior entropy therefore converges locally uniformly. A count over a shorter observation duration can be obtained by independent thinning of the count over a longer duration. Given the total count, the thinned count is conditionally independent of the source, establishing monotonicity of mutual information. Conditional on the shorter-duration count, the Gamma posterior remains nondegenerate, and the distribution of additional counts still depends on the source. The additional conditional mutual information is therefore strictly positive. The mutual information identity then implies strict decrease of the average posterior entropy.

## 5.4 Dirichlet Sources Under Multinomial Counting Encoding

Let $X \sim \text{Dirichlet}(\boldsymbol{\alpha})$, where $m \ge 2$, $\alpha_i > 0$, and $\alpha_0 = \sum_i \alpha_i$. For $t \ge 0$, draw a total count $N \sim \text{Pois}(t)$ independently of the source. Conditional on the input probability vector, make $N$ independent categorical draws and output their count vector:

$$Y|X = x, N = n \sim \text{Multinomial}(n, x), \qquad N = \textstyle\sum_{i=1}^{m} Y_i. \quad (5.16)$$

Equivalently, conditional on $X = x$, the components $Y_i \sim \text{Pois}(tx_i)$ are independent. The parameter $t$ is the expected total count, and the representation is a nonnegative integer vector. The total count is recoverable from the output and requires no separate description. By conjugacy [26], the induced output given $N = n$ is $\text{DirichletMultinomial}(n, \boldsymbol{\alpha})$, and the posterior is

$$X|Y = y \sim \text{Dirichlet}(\boldsymbol{\alpha} + y). \quad (5.17)$$

Define the entropy distortion of this channel by $E(t) = h(X|Y)$. Relative to Lebesgue measure on the free simplex coordinates $x_1, \dots, x_{m-1}$, the Dirichlet differential entropy is

$$h_D(\boldsymbol{\alpha}) = \log B(\boldsymbol{\alpha}) + (\alpha_0 - m)\psi(\alpha_0) - \textstyle\sum_{i=1}^{m}(\alpha_i - 1)\,\psi(\alpha_i), \quad (5.18)$$

where $B(\boldsymbol{\alpha}) = \prod_i \Gamma(\alpha_i)/\Gamma(\alpha_0)$. To compute $E(t)$, introduce the average posterior entropy at a fixed total count:

$$E_n := \mathbb{E}_{Y|N=n}[h_D(\boldsymbol{\alpha} + Y)], \qquad n = 0,1,\dots. \quad (5.19)$$

The expectation is taken under the conditional Dirichlet–multinomial output law. The quantity $E_n$ is an intermediate value independent of $t$. Conditioning on the total count yields

$$E(t) = e^{-t}\textstyle\sum_{n=0}^{\infty}\frac{t^n}{n!}E_n. \quad (5.20)$$

Hence this construction gives the continuous rate–entropy parametrization

$$R_e\big(E(t)\big) = h_D(\boldsymbol{\alpha}) - E(t), \qquad t \ge 0. \quad (5.21)$$

For a fixed number of draws, the count vector is a sufficient statistic of the categorical sequence for the source. Every additional draw supplies positive conditional mutual information, so $E_{n+1} < E_n$. Bounds on the coordinate posterior variances and subadditivity of differential entropy give

$$E_n \le \frac{m-1}{2}\log\frac{2\pi e}{4(\alpha_0+n+1)} \to -\infty. \quad (5.22)$$

The posterior entropy formula also gives $|E_n| \le C[1 + n\log(n+1)]$, where $C$ depends only on the source parameters. Thus (5.20) converges locally uniformly and can be differentiated term by term, yielding $E'(t) = \mathbb{E}[E_{N+1} - E_N] < 0$. Together with $E(0) = h_D(\boldsymbol{\alpha})$ and $E(t) \to -\infty$, this shows that the parametrization covers every finite $E \le h_D(\boldsymbol{\alpha})$.

For $m = 2$, let $P = X_1$, so that $P \sim \text{Beta}(\alpha_1, \alpha_2)$. Given $N = n$, the channel is $Y_1|P, N = n \sim \text{Binomial}(n, P)$, with posterior

$$P|Y = y \sim \mathrm{Beta}(\alpha_1 + y_1, \alpha_2 + y_2). \quad (5.23)$$

Beta-source encoding by binomial counts with a random total is therefore the two-dimensional case of this construction. Its rate function is also given by (5.20)–(5.21).

For Gamma reciprocal encoding with multiplicative noise, the posterior structure identifies a fixed IS distortion and recovers the established exact rate–distortion function [13]. For Gamma and Dirichlet counting channels, the conjugate posteriors give rate–entropy parametrizations over the full finite entropy-budget range. The same Gamma source produces different budget–rate relations through the independent additive residual of Section 4 and the conjugate posteriors here, reflecting the roles of both the distortion criterion and the representation.

## 6 The Rate–Entropy Coding Theorem and Its Constructive Proof

Sections 4 and 5 construct single-letter information rates through compatible posteriors. This section gives explicit block codes for entropy budgets: the encoder partitions the source space and sends the cell index, and the decoder forms the posterior within that cell. The construction provides an operational interpretation of the relation between rate and remaining uncertainty.

### 6.1 Block Codes and Entropy Distortion

Let $X^n = (X_1, \dots, X_n)$ be an independent and identically distributed source block. A deterministic length-$n$ encoder produces a message $Y^{(n)} = f_n(X^n)$ from an alphabet of size $M_n$. Define the rate and per-symbol block entropy distortion by

$$R_n = \frac{1}{n}\log M_n, \qquad E_n^{\mathrm{code}} = \frac{1}{n} H_{\mu^{\otimes n}}(X^n | Y^{(n)}). \quad (6.1)$$

The reference measure follows Section 2: counting measure for discrete sources and Lebesgue measure for continuous sources. A pair $(R, E)$ is achievable if there is a sequence of codes with $\limsup_{n\to\infty} R_n \le R$ and $\limsup_{n\to\infty} E_n^{\mathrm{code}} \le E$. The infimum of achievable rates is denoted by $R_e^{\mathrm{op}}(E)$. The message alphabet may depend on the blocklength and target entropy budget.

The message determines a block posterior. Under block logarithmic loss, let $q_n\left(\cdot\,|Y^{(n)}\right)$ be the decoder density and $Q_n\left(\cdot\,|Y^{(n)}\right)$ its distribution. Whenever the terms are well defined,

$$\frac{1}{n}\mathbb{E}[-\log q_n(X^n | Y^{(n)})] = E_n^{\mathrm{code}} + \frac{1}{n}\mathbb{E} D_{\mathrm{KL}}(P_{X^n|Y^{(n)}} \| Q_n(\cdot\,|Y^{(n)})). \quad (6.2)$$

The true block posterior minimizes expected block logarithmic loss. Definition (6.1) retains posterior dependence within a block. The average logarithmic loss obtained from the individual posterior marginals is $n^{-1}\sum_i H_\mu\left(X_i|Y^{(n)}\right)$, which is generally no smaller than (6.1). The operational rate function below uses block entropy distortion.

A deterministic encoder partitions the source block space into measurable message regions $B_j$ and sends the index of the region containing the source block. For a region with $P_{X^n}(B_j) > 0$, the decoder posterior is

$$K_n(x^n | j) = \frac{p_X^{\otimes n}(x^n)\mathbf{1}_{B_j}(x^n)}{P_{X^n}(B_j)}, \qquad P_{Y^{(n)}}(j) = P_{X^n}(B_j). \quad (6.3)$$

Zero-probability regions may be merged into any other region. The mixture identity $\sum_j P_{Y^{(n)}}(j) K_n(x^n | j) = p_X^{\otimes n}(x^n)$ shows that the encoding partition directly supplies a marginally compatible backward channel.

The discrete construction uses the weakly typical set. For an alphabet of size $q < \infty$, take $\varepsilon_n = n^{-1/4}$ and define

$$T_n = \left\{x^n : p_X^{\otimes n}(x^n) > 0,\ \left|-\frac{1}{n}\log p_X^{\otimes n}(x^n) - H(X)\right| \le \varepsilon_n\right\}. \quad (6.4)$$

Let $\eta_n = P_{X^n}(T_n^c)$. On a finite alphabet, the self-information has finite variance. Chebyshev's inequality gives $\eta_n \le \mathrm{Var}[-\log p_X(X)]/\sqrt{n} \to 0$, and the typical-set definition gives $|T_n| \le e^{n[H(X)+\varepsilon_n]}$.

## 6.2 The Rate–Entropy Coding Theorem and Proof

**Theorem 3 (Rate–entropy coding theorem).**

Let $X^n$ be an independent and identically distributed source block, with block entropy distortion as in Section 6.1. If either (i) the source has a finite alphabet and $E \ge 0$, or (ii) $X \in \mathbb{R}^m$ is absolutely continuous, $\int p_X(x)|\log p_X(x)|\,dx < \infty$, and $E$ is any finite real number, then

$$R_e^{\mathrm{op}}(E) = R_e(E) = [H_\mu(X) - E]^+. \quad (6.5)$$

The discrete case is achieved asymptotically by deterministic grouping of typical sequences, and the continuous case by per-sample equal-probability partitions. Both constructions use the decoder posterior in (6.3).

Proof. Converse. Since $Y^{(n)}$ is a deterministic function of $X^n$, we have $I(X^n; Y^{(n)}) = H(Y^{(n)})$. Hence

$$E_n^{\mathrm{code}} = H_\mu(X) - \frac{1}{n}H(Y^{(n)}) \ge H_\mu(X) - \frac{1}{n}\log M_n = H_\mu(X) - R_n. \quad (6.6)$$

Taking limits and using nonnegativity of rate gives $R_e^{\mathrm{op}}(E) \ge \left[H_\mu(X) - E\right]^+$.

**Achievability for discrete sources.**

If $E \ge H(X)$, a constant message suffices. Suppose $0 \le E < H(X)$, and use the typical set and probability bound from Section 6.1.

Fix an ordering of the alphabet and list the sequences in $T_n$ lexicographically. Divide this list into groups of at most $L_n = \lfloor e^{nE} \rfloor$ sequences, and place all atypical sequences in one additional group. The encoder sends the group index, and the decoder uses the within-group posterior (6.3). The number of messages satisfies

$$M_n \le \left\lceil \frac{|T_n|}{L_n} \right\rceil + 1 \le 2e^{n[H(X)-E+\varepsilon_n]} + 2. \quad (6.7)$$

A typical group has at most $L_n$ elements and therefore posterior entropy at most $\log L_n$. The atypical group has at most $q^n$ elements. Thus

$$E_n^{\mathrm{code}} \le (1-\eta_n)\frac{\log L_n}{n} + \eta_n \log q \le E + \eta_n \log q. \quad (6.8)$$

Equations (6.7)–(6.8) imply

$$\limsup_{n\to\infty} R_n \le H(X) - E, \qquad \limsup_{n\to\infty} E_n^{\mathrm{code}} \le E. \quad (6.9)$$

The construction includes $E = 0$: each typical sequence forms a singleton group, while the atypical group contributes asymptotically zero per-symbol entropy.

**Achievability for continuous sources.**

If $E \ge h(X)$, use a constant message. Otherwise, let $r_0 = h(X) - E > 0$, $b = \lfloor e^{r_0} \rfloor$, $\alpha = [r_0 - \log b]/[\log(b+1) - \log b]$, and $k_n = \lceil n\alpha \rceil$. Partition each of the first $k_n$ samples into $b+1$ equal-probability cells and each remaining sample into $b$ such cells. Denote the number of cells for sample $i$ by $m_i$. Let $F_1$ be the marginal distribution function of the first coordinate of a source vector. Absolute continuity and the probability integral transform [27] give

$$V_i = F_1(X_{i,1}), \qquad V_1, \ldots, V_n \sim \text{i.i.d.Unif}(0,1). \quad (6.10)$$

Divide the unit interval for each $V_i$ into $m_i$ equal subintervals. The encoder sends the vector of cell indices:

$$J_i = \min\{m_i, 1 + \lfloor m_i F_1(X_{i,1}) \rfloor\}, \quad Y^{(n)} = (J_1, \ldots, J_n), \quad M_n = \prod_{i=1}^n m_i. \quad (6.11)$$

Let $A_{i,j}$ be cell $j$ for sample $i$. The cell indices are independent and individually uniform. For message $j^n = (j_1, \ldots, j_n)$, the decoder posterior is

$$K_n(x^n \mid j^n) = \prod_{i=1}^n \left[m_i p_X(x_i) \mathbf{1}_{A_{i,j_i}}(x_i)\right]. \quad (6.12)$$

Each factor is normalized and has integrable log density under its posterior law. Mixing over messages recovers the independent source distribution. The samples remain conditionally independent given the message, so block conditional entropy is the sum of the individual conditional entropies. Using $H(J_i) = \log m_i$ gives

$$E_n^{code} = \frac{1}{n}\sum_{i=1}^n h\left(X_i \mid Y^{(n)}\right) = h(X) - \frac{1}{n}\sum_{i=1}^n \log m_i = h(X) - R_n. \quad (6.13)$$

Since $R_n = \log b + (k_n/n)\log[(b+1)/b]$ and $0 \le k_n/n - \alpha < 1/n$,

$$E_n^{code} \le E, \qquad 0 \le R_n - [h(X) - E] \le \frac{1}{n}\log\frac{b+1}{b} \longrightarrow 0. \quad (6.14)$$

This achieves the required rate for continuous sources. Together with the converse and Theorem 1, it proves (6.5).

The proof specifies the encoding regions, transmitted messages, and decoder posteriors directly. For discrete sources, group cardinalities control posterior entropy. For continuous sources, equal-probability cells make the message entropy equal to the logarithm of the message alphabet size. In comparison with random-covering proofs for general rate–distortion coding [2], [3], the entropy objective can be evaluated directly from group probabilities and within-group posteriors. No jointly typical sequence pairs or random-codebook covering-failure bounds are needed.

The linear rate function under discrete logarithmic loss is established in [16]–[18]; for its connection to list decoding, see [16], [18]. The construction here specifies an encoding map and decoder posterior for a block entropy budget. The continuous construction allocates cells separately to each sample and achieves both the block entropy target and the average individual conditional entropy target. Log-density integrability suffices, including for Cauchy sources without finite second moments. The discrete grouping construction is evaluated under the block entropy criterion of Section 6.1.

These codes assume a known source distribution. Their implementation complexity depends on typical-sequence enumeration or evaluation of the source distribution function. For a fixed physical distortion, the single-letter relation from Section 2 remains

$$R_d(D) = H_\mu(X) - E_d(D) = R_e(E_d(D)). \quad (6.15)$$

Under the assumptions of the classical rate–distortion coding theorem, (6.15) gives the corresponding minimum asymptotic rate [1]–[3]. Sections 4 and 5 supply candidate or optimal test channels. Operational achievability under a specified physical cost additionally requires control of that cost and may use the classical covering argument [28]. The partition construction in this section directly achieves the entropy budget.

## 7 Conclusions and Future Directions

We have developed a framework for constructing analytic rate–distortion functions based on marginal compatibility and attainment of maximum conditional entropy. A compatible posterior realizes the prescribed source distribution; its parameter structure identifies candidate fixed distortion functions; and the maximum conditional entropy conditions certify optimality. Convolution decompositions and conjugate posteriors provide two ways to carry out this construction. These paths can describe the same optimal joint distribution, as illustrated by the Gamma–IS example. Building on the authors' earlier rate–entropy formulation, we have also established attainability of general entropy levels and given a constructive coding proof for finite-alphabet sources and absolutely continuous sources with integrable log densities. The continuous case requires no finite second moment. Different physical distortion budgets determine different maximum conditional entropies, while the rate–entropy relation provides a common expression for the required information rate.

The framework recovers or explains established results for Gaussian, Laplace, exponential, Pareto, and Gamma sources under their respective distortion criteria. It also yields a closed-form parametric rate–distortion expression for zeta sources under divisibility-constrained logarithmic multiplicative distortion, an exact expression in terms of the source entropy over a specified interval for negative binomial sources under one-sided counting distortion, and an exact interval for Gamma sources under one-sided mean error. The geometric special case covers the full distortion range. The equivalent residual-count parametrization of a Poisson entropy budget and the rate–entropy parametrizations of Gamma and Dirichlet counting channels further connect model parameters with remaining uncertainty. Together, these results show how source structure can support both compatible-channel construction and the selection of distortion measures, with optimality verified under explicit constraints and parameter ranges.

Further theoretical work should characterize when a posterior family admits a fixed-distortion representation, determine the solvability of the associated marginal mixture equations, and extend the exact results to distortion intervals not covered by the present decompositions. The rate-gap identities may also support approximate constructions with quantifiable departures from optimality. These questions would broaden the classes of sources and costs accessible to analytic construction.

On the application side, distortion measures suggested by statistical structure should be assessed jointly with task requirements in compression and representation learning. IS distortion provides a criterion for relative errors in positive-valued features. Logarithmic multiplicative costs suggest directions for compressing hierarchical counts and statistical features of language. Conditional entropy and its physical-parameter coordinates offer ways to quantify uncertainty remaining in a representation. Relating these budgets to performance in image, video, and language tasks, and developing practical encoders that meet them, would connect the construction theory to distortion design and coding optimization.

## Appendix Summary of Distortion Criteria and Rate Functions

Table 1 collects the constraints, formulas, validity ranges, and provenance of the results. Expected-cost criteria and entropy-based budgets are listed separately. Entropy width, entropy scale, and equivalent residual count are monotone coordinates of a conditional entropy budget. Cauchy strength is a functional of the error distribution. All logarithms are natural.

Write $[u]^+ = \max\{u, 0\}$, $H_G(D) = (D+1)\log(D+1) - D\log D$, $H_P(t) = H(\mathrm{Pois}(t))$, and $H_t = H(\mathrm{zeta}(t))$. The functions $\kappa, J$, and $h_D$ are defined in the main text. In the error criteria below, $\hat{X} = g(Y)$ and $Z = X - \hat{X}$.

| Source or model | Distortion criterion | Rate function and range | Provenance |
|---|---|---|---|
| General sources | $H_\mu(X\|Y) \le E$ | $R_e(E) = H_\mu(X) - E$ on the attainable linear range; zero for $E \ge H_\mu(X)$. | Discrete relation [16]–[18]; continuous formulation [4]; general attainability in Theorem 1. |
| Scalar Gaussian | $\mathbb{E}(X-\hat{X})^2 \le D$ | $R(D) = \frac{1}{2}[\log(\sigma^2/D)]^+$, $D > 0$. | Classical. |
| Vector Gaussian matrix budget | $\mathbb{E}ZZ^T \preccurlyeq K$ | $R(K) = \frac{1}{2}\log\frac{\det\Sigma}{\det K}$, $0 \prec K \preccurlyeq \Sigma$. | Classical [2], [3]. |
| Vector Gaussian determinant budget | $\det\mathbb{E}ZZ^T \le D$ | $R(D) = \frac{1}{2}[\log(\det\Sigma/D)]^+$, $D > 0$. | Gaussian tight case of [11, Theorem 11], with a reparametrized budget. |
| Vector Gaussian entropy budget | $h(X\|Y) \le E$ | $R_e(E) = h(X) - E$; $E = \frac{1}{2}\log[(2\pi e)^m \det K]$, $0 \prec K \preccurlyeq \Sigma$. | Authors' earlier work [4]. |
| Laplace | $\mathbb{E}\|X-\hat{X}\| \le D$ | $R(D) = [\log(a/D)]^+$, $D > 0$. | Classical. |
| Uniform | $\exp\{h(X\|Y)\} \le D$ | $R(D) = [\log(a/D)]^+$, $D > 0$. Equal-width independent residuals realize $D = a/M$ for positive integers $M$. | Entropy-width relation [4]; other entropy levels by Theorem 1. |
| Cauchy | $e^{h(X\|Y)}/(4\pi) \le D$; separately, $s_C(Z) \le D$. | $R(D) = [\log(a/D)]^+$, $D > 0$, under either stated criterion. | Entropy scale [4]; Cauchy strength [22]. |
| Gamma | $\mathbb{E}(X-\hat{X}) \le D$, $0 \le \hat{X} \le X$. | $R(D) = \kappa(r) - 1 + \log(\theta/D)$, $r \ge 1$, $0 < D \le \theta$. Zero for $D \ge r\theta$. | Unified derivation here; exponential special case [6]. |
| Negative binomial and geometric | $\mathbb{E}(X-\hat{X}) \le D$, $\hat{X} \in \{0, \dots, X\}$. | $R(D) = H(X) - H_G(D)$, $0 \le D \le D_*$. Zero for $D \ge r(1-p)/p$. | Derived here; exact expression in terms of the source entropy. Geometric case covers the full range. |
| Poisson | $H_P^{-1}(H(X\|Y)) \le D$ | $R(D) = H_P(\lambda) - H_P(D)$, $0 \le D \le \lambda$; zero above $\lambda$. | Reparametrization of the entropy budget and attaining decomposition here. |

| Source or model | Distortion criterion | Rate function and range | Provenance |
|---|---|---|---|
| Uniform on a finite group | $d(x,\hat{x}) = \mathbf{1}\{x \neq \hat{x}\}$ | $R(D) = \log q - h_b(D) - D\log(q-1)$, $0 \leq D \leq 1 - 1/q$; zero above. | Classical. |
| Pareto | $d(x,\hat{x}) = \log(x/\hat{x})$, $1 \leq \hat{x} \leq x$. | $R(D) = [\log(1/(\alpha D))]^+$, $D > 0$. | Equivalent to [6] under a logarithmic transformation; unified derivation here. |
| Zeta | $d(x,\hat{x}) = \log(x/\hat{x})$, $\hat{x} \mid x$. | $D(t) = -\zeta'(t)/\zeta(t)$, $R\big(D(t)\big) = H_s - H_t$, $t \geq s > 1$. Zero for $D \geq D(s)$; $R(0) = H_s$. | Closed-form parametric expression derived here. |
| Gamma reciprocal multiplicative noise | $d(x,\hat{x}) = x/\hat{x} - \log(x/\hat{x}) - 1$ | $D(k) = \log k - \psi(k)$, $R\big(D(k)\big) = J(r) - J(k)$, $k \geq r$. Zero for $D \geq D(r)$. | Existing result [13, Proposition 3]; unified derivation here. |
| Gamma counting channel | $h(X\|Y) \leq E$; $Y = N(tX)/t$, $t > 0$. | $E(t) = \mathbb{E}\kappa(r + tY) + \log\frac{\theta}{1+\theta t}$, $R_e\big(E(t)\big) = h(X) - E(t)$. | Conjugacy [25]; rate–entropy parametrization and range here. |
| Dirichlet and Beta counting channels | $h(X\|Y) \leq E$; Poissonized multinomial counts. | $E(t) = e^{-t}\sum_{n=0}^{\infty}\frac{t^n}{n!}E_n$, $R_e\big(E(t)\big) = h_D(\alpha) - E(t)$, $t \geq 0$. | Conjugacy [26]; rate–entropy parametrization and range here. |

Here $h_b(u) = -u\log u - (1-u)\log(1-u)$, with $0\log 0 = 0$. For the negative binomial source, $D_* = (1-p)/p$ when $r \geq 1$, and $D_* = r(1-p)/[1 - r(1-p)]$ when $0 < r < 1$. The expression does not determine the intermediate distortion interval when $r \neq 1$. The source entropy can be evaluated by its convergent series.

Cauchy strength is $s_C(Z) = \inf\{b > 0: \mathbb{E}\log[1 + (Z/b)^2] \leq \log 4\}$. Scale parameters are positive; the negative binomial parameter satisfies $0 < p < 1$, and Dirichlet parameters are positive. The Gamma one-sided formula requires $r \geq 1$, whereas the Gamma IS and counting formulas allow every $r > 0$. The finite-group size satisfies $q \geq 2$. Costs are $+\infty$ outside the stated admissible error domains.

The counting channels use a constant output at $t = 0$ and cover all finite entropy levels below the source entropy as $t$ varies. In the Dirichlet formula, $E_n$ is the conditional entropy averaged over a fixed total count $n$; $E(t)$ is the final Poissonized budget. For continuous sources, the entropy range is finite $E \leq h(X)$; for discrete sources, it is $0 \leq E \leq H(X)$.